\documentclass{elsart}

\usepackage{natbib}

\usepackage{epsfig}

\usepackage{amssymb}

\begin{document}

\begin{frontmatter}



\title{The Crab pulsar seen with Aqueye at Asiago Cima Ekar Observatory}


\author[a1]{L. Zampieri\corauthref{cor}},
\corauth[cor]{Corresponding author}
\ead{luca.zampieri@oapd.inaf.it}
\author[a2,a1]{C. German\`a},
\author[a2]{C. Barbieri},
\author[a3,a4]{G. Naletto},
\author[a5]{A. \v{C}ade\v{z}},
\author[a3,a2]{I. Capraro},
\author[a6]{A. Di Paola},
\author[a2]{C. Facchinetti},
\author[a3,a2]{T. Occhipinti},
\author[a5]{D. Ponikvar},
\author[a2,a3]{E. Verroi} and
\author[a7,a2]{P. Zoccarato}
\address[a1]{INAF-Astronomical Observatory of Padova, Vicolo dell'Osservatorio 5, 35122 Padova, Italy}
\address[a2]{Astronomy Department, University of Padova, Vicolo dell'Osservatorio 2, 35122 Padova, Italy}
\address[a3]{Department of Information Engineering, University of Padova, Via Gradenigo 6/B, 35131 Padova, Italy}
\address[a4]{CNR/INFM/LUXOR, via Gradenigo 6/B, 35131 Padova, Italy}
\address[a5]{Faculty of Mathematics and Physics, University of Ljubljana, Jadranska 19, 1000 Ljubljana, Slovenia}
\address[a6]{INAF-Astronomical Observatory of Rome, Via frascati 33, 00040 Monteporzio Catone (Roma), Italy}
\address[a7]{Interdipartimental Center of Studies and Activities for Space (CISAS) ``G. Colombo'', University of Padova, Via Venezia 15, 35131 Padova, Italy}

\begin{abstract}

We are developing fast photon-counter instruments to study the rapid variability
of astrophysical sources by time tagging photon arrival times with
unprecedented accuracy, making use of a Rubidium clock and GPS receiver. The first realization of such optical 
photon-counters, dubbed Aqueye (the Asiago Quantum Eye), was mounted in 2008 at 
the 182cm Copernicus Observatory in Asiago. Aqueye observed the Crab pulsar several 
times and collected data of extraordinary quality that allowed us to perform
accurate optical timing of the Crab pulsar and to study the pulse shape stability
on a timescale from days to years with an excellent definition. Our results reinforce the 
evidence for decadal stability of the inclination angle between the spin and 
magnetic axis of the Crab pulsar. Future realizations of our instrument will 
make use of the Galileo Global Navigation Satellite System (GNSS) time signal.

\end{abstract}


\end{frontmatter}

\parindent=0.5 cm

\section{Introduction}

The pulsar in the Crab Nebula (PSR J0534+2200) is one of the best studied objects in the sky.
It is the brightest optical pulsar and the first to be detected in the 
optical band \citep{1969Natur.221..525C, 1969ApJ...155L.121L}. 
The pulse shape of the Crab pulsar was investigated in various photometric bands 
\citep[e.g.][]{1993ApJ...407..276P,1999AAS...194.5207L}.
It was found to be very stable, despite the secular decrease of 
luminosity \citep{1996A&A...314..849N,2009A&A...504..525S} and the presence of glitches and timing noise. 
Occasionally, small variations of the pulse shape have been observed 
\citep{2007arXiv0709.2580K}. Wavelength-dependent changes of pulsar properties have been 
reported also by \cite{2002ApJ...581..485F}. 

The first optical timing studies of the Crab pulsar were carried out soon after its
discovery \citep*{1970ApJ...161L.235N,1970Natur.228..445P,1971ApJ...166L..91H}, showing 
evidence of secular slowdown, noise and glitches \citep*{1972ApJ...175..217B}.
Besides a continuous monitoring of its timing behavior \citep*{1981A&AS...44....1L}, much attention
was devoted to the short term stability of the optical pulse shape \citep*{2007Ap&SS.308..595K}
and to the simultaneous absolute timing at radio and optical wavelengths \citep*{2006A&A...456..283O,2008A&A...488..271O}.
Most notably, \citet{2003Sci...301..493S} found that optical 
pulses synchronous with giant radio pulses (occasional powerful pulses that are a thousand
times brighter than average pulses) are 3\% brighter on average than those coincident 
with normal radio pulses. 

Short timescale (few minutes) modulations of the phase and amplitude of the optical light 
curve were investigated by \citet{1996A&A...306..443C}, finding evidence
for a 60 s modulation that was interpreted as the pulsar free precession period.
Although further measurements are needed in order to confirm this finding (see e. g. the 
different result obtained by \citealt{2000A&A...363..617G}), in principle it could be
used to constrain the pulsar moment of inertia and hence the equation of state of
nuclear matter. Theoretical models of a freely precessing neutron star were 
calculated and compared against pulsar observations by \cite{2001MNRAS.324..811J}.

It is widely accepted that, for young pulsars, optical emission is synchrotron radiation 
from relativistic particles spiraling around pulsar magnetic field lines. For middle-aged 
pulsars, in addition to the emission from the polar caps heated by impinging current, a 
significant contribution comes from thermal radiation from the cooling neutron 
star surface. The basic energy source is the pulsar
rotational energy, that is somehow transferred to low-frequency radiation and into 
accelerating charged particles. The major uncertainty is related to the acceleration 
mechanism of this relativistic wind. Very little is known about the acceleration site,
either at magnetic poles, at the surface or further out near the light cylinder, where
particles corotating with neutron star magnetic field lines reach the velocity of light.
Precise timing of pulsar light curves in different wavebands is a
powerful tool to constrain theories on the spatial distribution of various emission 
regions. While a time delay between different radio bands measures dispersion due to 
presence of electrons in the whole space between us and the pulsar, differences
in arrival times between the radio and optical or X-ray bands most naturally imply 
that the emission regions differ in position.
It has been shown that the main pulse and the interpulse are not aligned in time in the
radio, X-ray and Gamma-ray bands, but the high energy photons lead the radio ones
\citep{2003A&A...411L..31K,2004ApJ...605L.129R}. At optical and radio wavelengths
\cite{2006A&A...456..283O} and \cite{2008A&A...488..271O} recently performed simultaneous 
absolute timing measurements and found a $255\pm 21 \mu{\rm s}$ delay between the radio and 
the optical pulse (with the optical pulse leading the radio one). However, there does not
seem to be universal agreement on the value of this phase difference.

As it is bright and nearby and its timing properties are very accurately known,
the Crab pulsar is an ideal target for testing photon-counting instruments in all bands of
electromagnetic spectrum. It is used as a standard candle to calibrate both the flux and 
timing accuracy of optical and high-energy instruments. Thus, we decided to
test our novel photon-counter AquEYE \citep[the Asiago Quantum Eye;][]{2009JMOp...56..261B} 
by performing a sequence of observations of Crab pulsar. Here we report some preliminary
results from these observations.

\section{AquEYE}

AquEYE stands for Asiago Quantum Eye and is a very fast single photon-counter
based on avalanche photodiodes operated in Geiger mode (SPADs).
The design of AquEYE follows that of QuantEYE 
\citep[the Quantum EYE;][]{2005astro.ph.11027D,2009JMOp...56..261B}, an instrument 
conceived for studying second order correlations in photon streams from 
astrophysical sources, down to the quantum limit. QuantEYE was designed
for future very large collecting area telescopes, such as the extremely large 
European telescope of ESO (E-ELT).
AquEYE is mounted at the 182cm Copernicus telescope at Cima Ekar in Asiago.
It can record and store arrival times of all detected photons with an absolute precision 
(referred to UTC) better than 500 ps for hour-long observing sessions 
\citet{2010zoc}. To our knowledge, 
this instrument provides the most accurate arrival times ever achieved in the optical band.


Each signal from the SPADs is timetagged by a Time to Digital Converter (TDC)
board (produced by Costruzioni Apparecchiature Elettroniche Nucleari, Italy),
that makes use of an external Rubidium oscillator as external reference frequency.
This clock is extremely accurate on short term, but has a drift on long periods. 
To remove this drift, a pulse-per-second (PPS) signal from a GPS unit is given 
in input to the TDC board and time tagged together with the other recorded events.
Then a post-process analysis of the collected PPS allows to determine and remove 
the Rubidium drift. 
Future realizations of our instrument will make use of the 
Galileo GNSS time signal for improving the long term accuracy of our clock.

\begin{table}
    \caption{Log of the observations of the Crab pulsar performed 
      with Aqueye mounted at the 182cm Copernicus telescope in Asiago on the
      nights of October 10-12, 2008. The start time 
      of the observations is the GPS integer second, accurate to better than 
      approximately $\pm$30 nanoseconds. The second column is the MJD at the
      beginning of the observation corrected at the solar system barycenter.}
    \begin{tabular}{ccc}
      \hline
            Starting time &  MJD  &  Duration  \\
            (UTC)         &       &  (s)       \\      
      \hline
            Oct 10, 23:45:14  & 54749.993034053951089  &  1078  \\
            Oct 11, 01:45:44  & 54750.076722217817146  &  1797  \\
            Oct 11, 02:23:07  & 54750.102685217842115  &  1631  \\
            Oct 11, 23:25:08  & 54750.979164409017692  &  3597  \\
            Oct 12, 23:13:57  & 54751.971486747642043  &  3998  \\            
      \hline
      \end{tabular}
    \label{tab1}
\end{table}

\section{Data analysis}

During 2008 the Crab pulsar was observed several times with AquEYE.
Here we report on results of our analysis of some October 2008
observations, which are of particularly good quality 
in terms of sky and seeing conditions. For a log of observations, see Table~\ref{tab1}.
Arrival times of detected photons are corrected to the Solar System
barycenter using the software TEMPO2\footnote{http://www.atnf.csiro.au/research/pulsar/ppta/tempo2}
\citep{2006MNRAS.369..655H, 2006MNRAS.372.1549E}. The barycentric corrected arrival times were then 
processed with the timing analysis software XRONOS\footnote{http://xronos.gsfc.nasa.gov/} (v. 5.21). 
Following the default convention adopted in TEMPO2, in this analysis we use the barycentric 
coordinate time (TCB) as system of time.

\section{Results}

%

\begin{table}
      \caption{Rotational period of the Crab pulsar measured with Aqueye in 2008.
      The second column is the MJD at mid observation corrected at the solar system barycenter.}
      \begin{tabular}{ccc}
            \hline
            Starting time  &  MJD at mid observation  &  Period \\
            (UTC)          &                          &  (s)    \\
            \hline
            Oct 10, 23:45:14  &  54749.999284623447164  &  0.0336216392  \\
            Oct 11, 01:45:44  &  54750.087139828703616  &  0.0336216424  \\
            Oct 11, 02:23:07  &  54750.113797334172066  &  0.0336216433  \\
            Oct 11, 23:25:08  &  54750.999988051155317  &  0.0336216755  \\
            Oct 12, 23:13:57  &  54751.994405470224461  &  0.0336217117 \\
            \hline
      \end{tabular}
    \label{tab2}
\end{table}            

We searched for periodicities in the barycentric corrected time series 
of the observations by folding the data over a range of periods and by looking for a
maximum chi-square as a function of period (XRONOS task {\tt efsearch}).
We binned the data using 1000 bins per period and a resolution between contiguous
periods in the search of $10^{-10}$ s. 
The measured periods are reported in Table~\ref{tab2}.
The start time is the GPS integer second, accurate to better than approximately $\pm$30 nanoseconds.
For the observation performed on 11 October 2008, starting at 01:45:44 UTC,
the period is $P=0.0336216424$ s. For comparison, the period at 
mid observation obtained interpolating the radio Jodrell Bank 
Crab ephemerides is $P=0.0336216423$ s (after transforming from
the Barycentric Dynamical Time to the Barycentric Coordinate Time used in TEMPO2).
The difference with respect to our measured period is 0.1 ns. 
We can consider this as an estimate of our present uncertainty on the period measurement 
in a single observation. Our timing accuracy is comparable to the rotational period change
induced by the pulsar spin-down during the observation. 


An estimate of the period derivative can be obtained by fitting
directly the measurements reported in Table~\ref{tab2} as a function of time with a 
first order polynomial. We refer the period
measurements to mid observation. At the barycentric corrected time
$t_0=54749.0$ (MJD) the period and period derivative are 
$P=0.033621602861\pm 8.7 \times 10^{-11}$ s and
${\dot P}=(4.2061\pm 0.0056)\times 10^{-13}$ s/s (2$\sigma$ errors), 
within 0.06 ns and 0.005\%, respectively, from the Jodrell Bank Crab ephemeris
(after correcting from the Barycentric Dynamical Time to the Barycentric Coordinate Time).

\begin{figure}
\begin{center}
\includegraphics*[angle=0,width=12cm]{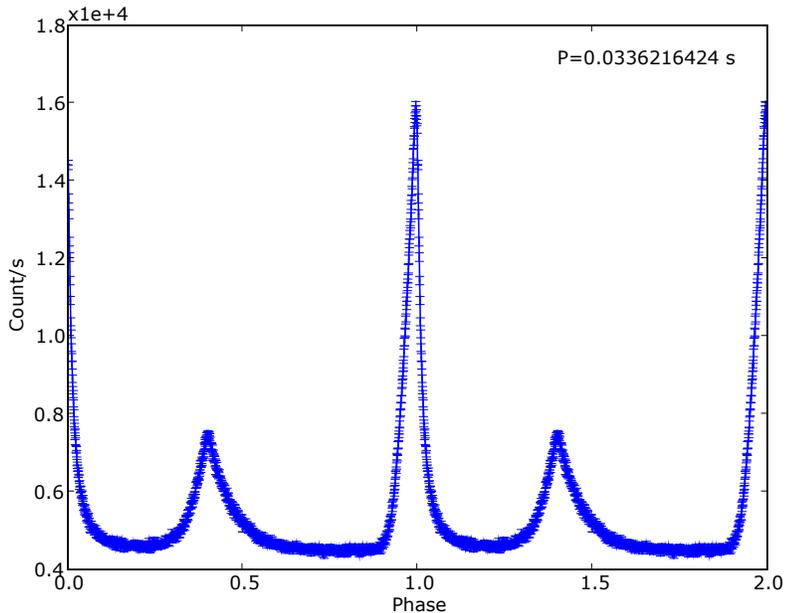}
\end{center}
\caption{Folded light curve over the average period of the Crab pulsar for the AquEYE 
observation performed on 11 October 2008 (01:45:44 UTC). The bin time of the light curve is $3.362\times 10^{-5}$ s.
For the sake of clarity two rotations of the neutron star are shown.}
\label{fig4}
\end{figure}

\begin{figure}
\begin{center}
\includegraphics*[angle=0,width=14cm]{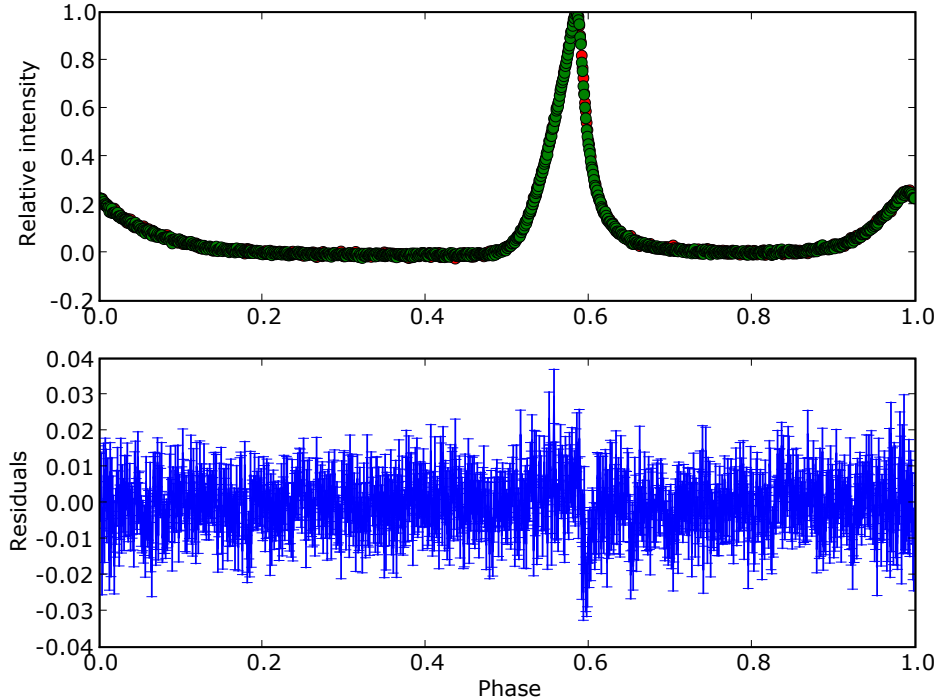}
\end{center}
\caption{{\it Upper panel}: AquEYE folded light curve 
of the Crab pulsar on 11 October 2008 ({\it green}; observation starting at 
01:45:44 UTC) and on 12 October ({\it red}; observation starting at 23:13:57 UTC). 
A shift in phase and a stretch in amplitude was applied to the Oct 11 data. 
{\it Bottom panel}: fractional residuals. The reduced $\chi^2$ of the difference 
of the two curves is 1.05.}
\label{fig4b}
\end{figure}

\begin{figure}
\begin{center}
\includegraphics*[angle=0,width=14cm]{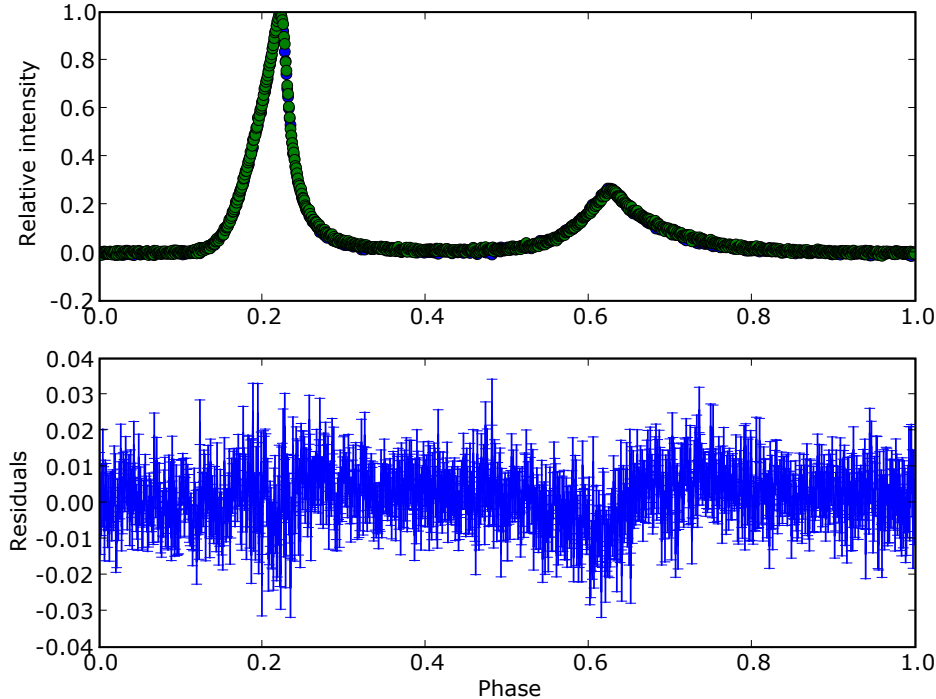}
\end{center}
\caption{{\it Upper panel}: AquEYE folded light curve of the Crab pulsar ({\it
green}; 11 October 2008, 01:45:44 UTC) and the $B$ band light curve obtained in 1994 
by \cite{2000NuPhS..80C1103B} ({\it blue}). A shift in phase and a stretch in amplitude
was applied to our data. {\it Bottom panel}: fractional residuals. The reduced $\chi^2$ of the difference
between the two curves is 1.3. A similar result is obtained by performing the comparison 
with the $R$ band light curve of \cite{2000NuPhS..80C1103B}.}
\label{fig5}
\end{figure}

In Figure~\ref{fig4} we show the light curve of the Crab pulsar folded over the 
average period for the 11 October 2008, 01:45:44 UTC. The folded curve has 1000 bins 
in phase, with a resolution of $\sim$33.6 $\mu$s, and contains data from 53,458 pulse periods. 
The double peak profile of the pulse has an excellent definition and its quality is 
comparable to that achieved with instrumentation mounted on larger area telescopes. 
In Figure~\ref{fig4b} this pulse profile is compared with that of the fifth observation
(Oct 12, 23:13:57 UTC), taken two nights after. The comparison is performed applying 
a shift in phase and a stretch in amplitude to one of the two curves, estimated 
using a $\chi^2$ minimization procedure. The two profiles are clearly overimposed
and no significant short term (days) variability is observed (apart, possibly, from a small 
difference in the trailing edge of the main peak, around phase 0.6, that will
deserve further investigation).
In Figure~\ref{fig5} we compare the Aqueye folded curve 
with that obtained in 1994 by \citet{2000NuPhS..80C1103B}, with a photomultiplier having 3.3 $\mu$s
resolution. The comparison is performed as before, by means of a $\chi^2$ minimization procedure.
In order to better compare the two curves, that of
\citet{2000NuPhS..80C1103B} was rebinned to 1000 bins in phase.
As shown in the plot of the residuals, the profiles are consistent and testify a remarkably 
stable behaviour, at the level of $\sim$1\% on a timescale of 14 years. There may be
some small systematic difference in the profile of the interpulse, but the
variation is within our present uncertainty. 
A similar pulse shape was also recorded in years 1999 
and 2003. We do not find evidence of the drastic change and variability of 
the light curve observed in 2005-2006 \citep{2007arXiv0709.2580K}.

\section{Conclusions}

We observed the Crab pulsar with a novel photo-counter, AquEYE, mounted at the 182cm 
Copernicus telescope in Asiago. The observations performed on October 10-12, 2008
were of particularly good quality in terms of sky and seeing conditions.
The counting statistics allowed us to determine the pulsar
rotational period and period derivative with great accuracy from only three
nights of data. 
A full statistical analysis of the time-of-arrival and phase of the main pulse
in comparison with those of the Jodrell Bank radio ephemerides
is presently under way \citep*{2010germ} and may possibly allow us to obtain an 
independent measurent of the delay between the radio and the optical pulse.

Apart from the changes visible only in 2005-2006 \citep{2007arXiv0709.2580K}, during the
last 14 years the folded light curve of the Crab pulsar was very stable, as we measured
a pulse shape essentially 
identical to that observed in 1994. Any variation is below the sum of the statistical 
errors of the two curves ($\simeq$1\%). Since 1994 the frequency of the pulsar
changed by more than 0.5\% (or, if the dipole radiation loss dominates, the rotational 
energy loss decreased by more than 2\%). Pulse shapes are typically quite stable, 
unless the pulsar undergoes frequent mode switching, in which case an abrupt change of the pulse with respect to 
the usual integrated pulse profile may occur at unpredictable times. Such behaviour is observed 
in some complex profile radio pulsars (mode switching pulsars; e. g. \citealt{1970Natur.228..752B, 1982ApJ...258..776B}). 
The steadiness and large duty cycle appear to be consistent 
with the outer gap accelerator model of pulsar emission \citep{2007ApJ...670..677T}.
Pulse profiles of Crab-like pulsars from the optical through the 
gamma-ray bands have been investigated within the framework of this model
(see e. g. \citealt{2009vent}). In particular, \cite{2007ApJ...670..677T}
find that differences in pulse shapes are caused mainly by: differences
in inclination angles ($\alpha$) between the spin axis and the magnetic axis; differences in viewing 
angles; differences in thickness of the emission region. For the Crab pulsar, they found
$\alpha \approx 50^0$. Our data therefore reinforce evidence for decadal stability 
of the inclination angle $\alpha$, the viewing angle
and the thickness of the emission region.


Using the experience gained with AquEYE,
a new, improved version of the instrument, dubbed IquEYE (Italian Quantum EYE), has 
already been built and mounted at the ESO NTT. Two runs performed in January and December 2009 permitted to test 
the instrument and acquire useful data which are now being reduced. 
We intend to search for possible modulations in the Crab pulsar induced by free precession. 
At the same time, we are planning simultaneous, high time
resolution observations in the radio, optical, X-ray and gamma-ray bands in order
to study the delays of the pulse shape among different energy bands.
For future realizations, the presently  available GPS signal  could be supplemented by 
those of the Galileo GNSS. Having more satellites with different characteristics 
could help to lower the uncertainties in the start time and improve both the long 
term stability and accuracy of our timing system.

\section*{Acknowledgments}

We would like to thank Andrea Possenti (INAF-Astronomical Observatory of Cagliari) 
for his help in using the TEMPO2 software.
We also thank Massimo Calvani (INAF-Astronomical Observatory of Padova) and Alessandro Patruno 
(Astronomical Institute, University of Amsterdam) for useful discussions.
We are grateful to Sergey V. Karpov for kindly providing us the files of 
the 1994 light curve of the Crab.
This work has been partly supported by The University of Padova, by the Italian Ministry 
of University MIUR through the program PRIN 2006, by the Program of Excellence 2006 
Fondazione CARIPARO, and by the Harrison project of the Galileo supervising authority.


\end{document}